\documentstyle[aps,prl,twocolumn,floats,graphicx]{revtex}

\begin{document}
\draft
\preprint{HEP/123-qed}
\wideabs{
\title{Vanishing of phase coherence in underdoped ${\bf Bi_{2}Sr_{2}CaCu_{2}O_{8+\delta}}$}
\author{J. Corson, R. Mallozzi and J. Orenstein}
\address{Materials Sciences Division, Lawrence Berkeley National Laboratory and Physics Department, University
of California, Berkeley, California 94720}
\author{J.N. Eckstein}
\address{Department of Physics, University of Illinois, Urbana, Illinois 61801}
\author{I. Bozovic}
\address{Oxxel GMBH, D-28359 Bremen, Germany}
\date{\today}
\maketitle
\begin{abstract}
While the binding of electrons into Cooper pairs is essential in forming the superconducting state, its remarkable properties -  zero resistance and perfect diamagnetism - require phase coherence among the pairs as well.  When coherence is lost at the transition temperature $T_c$, pairing remains, together with phase correlations which are finite in space and time.  In conventional metals Cooper pairs with short-range phase coherence survive no more than 1 K above $T_c$.  In underdoped high-$T_c$ cuprates, spectroscopic evidence for some form of pairing is found up to a temperature $T^{*}$ which is roughly 100 K above $T_c$\cite{1,2,3}.  How this pairing and Cooper pair formation are related is a central problem in high-$T_c$ superconductivity.  The nature of this relationship hinges on the extent to which phase correlations accompany pairing in the normal state\cite{4}.  Here we report measurements of high-frequency conductivity which track the phase-correlation time $\tau$ in the normal state of the $Bi_{2}Sr_{2}CaCu_{2}O_{8+\delta}$ family of underdoped cuprate superconductors.  Just above $T_c$, we find that $\tau$ reflects the motion of thermally generated topological defects in the phase, or vortices\cite{5,6}.  However, vortex proliferation reduces $\tau$ to a value indistinguishable from the lifetime of normal state electrons at 100 K, well below $T^{*}$. 
\end{abstract}

}
In layered superconductors like the cuprates, the density of paired electrons per layer, $N_{s}$, determines the phase stiffness, the energy cost to produce spatial variations in phase within a layer.  Superconductivity persists until the thermal energy $k_{B}T$ becomes comparable to the phase-stiffness energy $k_{B}T_{\Theta} \equiv N_{s}\hbar^{2}/m^{*}$, where $m^{*}$ is the carrier effective mass.  In conventional superconductors at low temperature $k_{B}T_{\Theta}$ is much greater than the pair binding energy.  In such materials the transition to the normal state is driven by Cooper pair unbinding, which reduces $N_{s}$ until $T_{\Theta}$ becomes comparable to $T$.  

\begin{figure}[h]
     \includegraphics[width=3in]{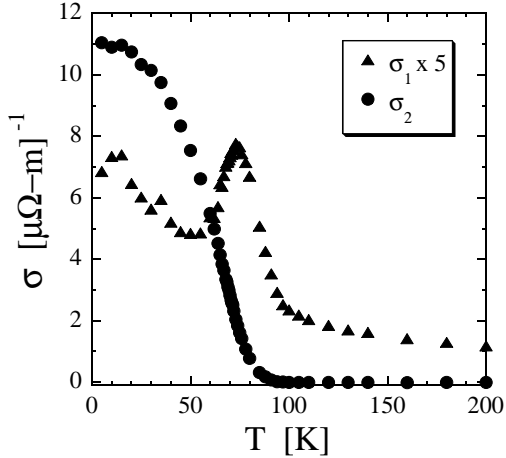}
\caption{The complex conductivity $\sigma$ measured at 100 GHz, as a function of the temperature $T$.  The real part, $\sigma_1$, is multiplied by 5 for ease of comparison with the imaginary part $\sigma_2$.  The real part is comprised of a peak near $T_{c}$ superposed on a smooth background.  We interpret the peak as the contribution from a partially coherent superfluid and the background as the contribution from quasiparticles.}
\label{fig:First}
\end{figure}

In copper oxide superconductors the conventional ordering of the binding and phase stiffness energies is reversed\cite{4}.  As their carrier concentration is modified, away from the optimal value for highest $T_{c}$ towards the Mott insulator (underdoping), this reversal becomes more profound; reduced phase stiffness is accompanied by larger binding energy\cite{2,7}.  Thus, in underdoped cuprates, electrons may remain tightly bound while long-range phase coherence vanishes at $T_{c}$\cite{4,8}.  A state with paired electrons and short-range phase correlations was proposed to account for the anomalous properties of cuprates above the transition temperature, between $T_{c}$ and $T^{*}$\cite{4}. 

This conjecture has led to the search for normal state remnants of the infinite DC conductivity and perfect diamagnetism which exist below $T_{c}$.  For the most part, these consequences of partial phase coherence have not been observed.  It can be argued, however, that if the correlation times are sufficiently short, the expected enhancement above normal state values may be unobservable by low-frequency probes.

The above argument motivates the use of high-frequency techniques to capture the short time-scale dynamics.  A powerful probe of this type is the complex frequency dependent conductivity $\sigma(\omega)$.  Below $T_{c}$, the conductivity measures the phase-stiffness energy directly: 
\begin{equation}
\sigma(\omega)=i\sigma_{Q}(k_{B}T_{\Theta}/\hbar\omega)
\label{eq:first}
\end{equation}
where $\sigma_{Q}\equiv e^{2}/\hbar d$ is the quantum conductivity of a stack of planar conductors with interlayer spacing $d$.  Equally important is the behavior of $\sigma(\omega)$ expected above $T_{c}$ in the presence of short-range phase correlations.  At low frequency, $\sigma$ will approach a real constant: the normal state DC conductivity.  At sufficiently high probing frequency, however, the phase-fluctuating state becomes indistinguishable from the superconducting one and $\sigma$ is expected to approach Eq. 1.  The crossover between these two limits occurs at a frequency $\Omega\equiv 1/\tau$, where $\tau$ is the phase-correlation time.

In this work, we report $\sigma(\omega)$ measured on a set of underdoped cuprate superconductors.  We use the technique of time-domain transmission spectroscopy to capture the linear response at high-frequency, from 100-600 GHz\cite{9}.  Direct measurement of electric field rather than intensity yields both the real and imaginary parts of $\sigma(\omega)$ without Kramers-Kronig analysis.  The samples are thin (40-65.6 nm) epitaxial films of underdoped $Bi_{2}Sr_{2}CaCu_{2}O_{8+\delta}$ (BSCCO) grown by atomic layer-by-layer molecular beam epitaxy\cite{10}.  Underdoping is achieved by either varying oxygen concentration or replacement of calcium by dysprosium.  Angle-resolved photoemission measurements a function of carrier concentration have been performed on BSCCO samples grown in the same chamber\cite{7}. 

Fig. 1 shows the real and imaginary parts of the conductivity, $\sigma_1$ and $\sigma_2$, at 100 GHz for an underdoped BSCCO film.  In this sample, the resistive $T_{c}$ was reduced to 74 K by replacement of 10 $\%$ of Ca by Dy.  The component related to dissipation, $\sigma 1$, has a peak centered near $T_c$ superposed on a rising background.  Notice that $\sigma_2$ becomes observable near 100 K, more than 25 K above $T_{c}$ and has no obvious onset temperature.  Both curves show that the transition from normal to superconducting state is broadened when viewed on a short time scale.  While partial phase coherence above $T_{c}$ can give rise to such effects, more conventional interpretations must be considered as well.  These include broadening due to static disorder, or Gaussian fluctuations, in which the pairing amplitude fluctuates about zero and the phase is undefined.   

To help distinguish these possibilities we examine the behavior of $\sigma_2$ for different frequencies.  For comparision with theory we convert $\sigma_2$ to the phase stiffness of an individual $CuO_{2}$ bilayer using Eq. 1, $k_{B}T_{\Theta}(\omega)\equiv \hbar \omega(\sigma_{2}/\sigma_{Q})$, where $\sigma_{Q}=1.58\times 10^{5} \Omega^{-1}m^{-1}$ for a spacing between bilayers of 1.54 nanometers.  Fig. 2 shows $T_{\Theta}(\omega)$ vs. $T$ on a semilog scale for the sample shown in Fig. 1 and for our most underdoped sample with $T_{c}$=33 K.  This plot identifies a crossover in the dynamics with increasing temperature.  At low $T$ the phase-stiffness is frequency independent.  With increasing temperature, $T_{\Theta}$ begins to fan out for the different frequencies in our bandwidth, with the lowest frequency data decreasing most rapidly.  All five samples that we have measured show the behavior described above.  Furthermore, we find that for all samples the value of $T_{\Theta}$ at the crossover, and the temperature at which it occurs, are related linearly.  The dashed line in Fig. 2 shows how well the crossover is described by the simple relation, $T_{\Theta}=(8/\pi)T$.  

\begin{figure}[h]
     \includegraphics[width=3in]{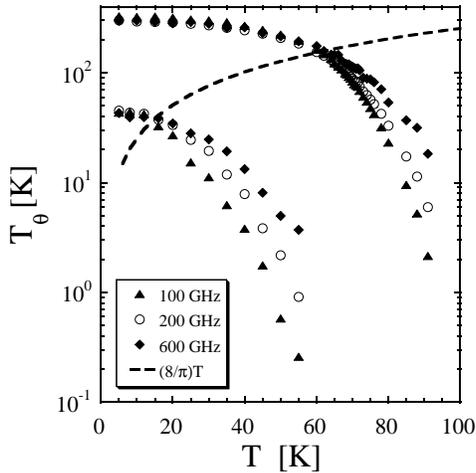}
\caption{The dynamic (frequency dependent) phase-stiffness temperature, $T_{\Theta}(\omega)$ as a function of temperature $T$.  Data are shown for two samples, one with $T_{c}$=33 K  (left side) and the other with $T_{c}$=74 K (right side).  The three curves for each sample correspond to measurement frequencies of 100, 200 and 600 GHz.  The family of curves identify a crossover from frequency independent to frequency dependent phase stiffness.  The dashed line shows that the crossover corresponds to the KTB condition for 2D melting, i.e. when the phase stiffness and the temperature are related by $T_{\Theta}=(8/\pi)T$.}
\label{fig:Second}
\end{figure}

The phase-stiffness dynamics shown in Fig. 2 suggest the relevance of the Kosterlitz-Thouless-Berezinskii (KTB) theory of 2D melting\cite{5,6}.  In the KTB picture, phase coherence is controlled by thermally-generated vortices.  In 2D each vortex is a point defect around which the phase changes by $2\pi$.  In the ordered state vortices bind into pairs of opposite vorticity.  The transition to the disordered state occurs when the first unbound vortices appear.  Their random thermal motion leads to a finite phase-correlation time.  Regardless of the microscopic details which distinguish one material from another, the transition is predicted to occur when the reduced temperature $T/T_{\Theta}$  reaches a universal value of $\pi/8$.  

\begin{figure}[h]
     \includegraphics[width=3in]{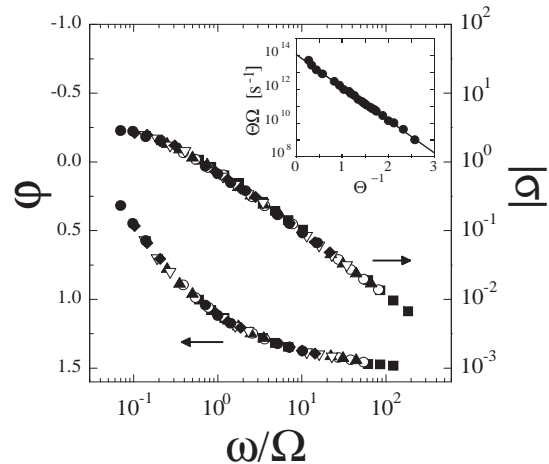}
\caption{Main panel shows the conductivity phase angle, $\phi=tan^{-1}(\sigma_{2}/\sigma_{1})$, and normalized conductivity magnitude $|\sigma|$, plotted as a function of the reduced frequency $\omega/\Omega(T)$.  In calculating $\phi$ we include only the fluctuation contribution to $\sigma_1$, which is obtained by subtracting the broad quasiparticle background from the peak seen in Fig. 1.  The scales for $\phi$, and $|\sigma|$, are shown on the left-hand, and right-hand axes, respectively.  Each plot is comprised of data measured at temperatures in the range from 64 to 91 K, and frequencies from 100 to 400 GHz.  Normalizing the frequency scale to the fluctuation frequency $\Omega(T)$ collapses the experimentally determined $\phi(\omega,T)$ to a single curve.  A similar data collapse as a function of $\omega/\Omega(T)$ is achieved when the conductivity magnitude is normalized by the quantity $\sigma_{Q}k_{B}T^{0}_{\Theta}/\hbar\Omega$.  The values of $T^{0}_{\Theta}(T)$ and $\Omega^{-1}(T)$ found by this scaling analysis are plotted in Fig. 4.  The inset shows a remarkable exponential relationship between the fluctuation frequency $\Omega$ and the reduced phase-stiffness temperature $\Theta\equiv T/T^{0}_{\Theta}$.  The graphs shows the $\Theta\Omega$ vs. $1/\Theta$ on a semilog scale.  The line fit through the data corresponds to the exponential relation, $\Omega=(\Omega_{0}/\Theta)exp(-2C/\Theta)$, with $C=2.23$ and $\Omega_{0}=1.14\times 10^{14} s^{-1}$.}
\label{fig:Third}
\end{figure}

The experimental signature of a KTB transition is the behavior of the dynamical phase stiffness.  At the critical temperature, $T_{KTB}$, the low-frequency $T_{\Theta}$ drops discontinously to zero from its value just below the transition, $(8/\pi)T_{KTB}$\cite{11}.  However, for $\omega>\Omega$, $T_{\Theta}$ approaches the "bare" phase-stiffness of the underlying superconductivity, $T_{\Theta}^{0}$, and varies smoothly through the transition.  At intermediate frequencies, $T_{\Theta}(\omega)$ interpolates smoothly between the these two limiting behaviors\cite{12}.

The behavior seen in Fig. 2 is consistent with KTB dynamics, if we identify the crossover with $T_{KTB}$ of an isolated bilayer.  Above $T_{KTB}$, the conductivity is predicted to scale according to\cite{13,14,15}:
\begin{equation}
\frac{\sigma(\omega)}{\sigma_{Q}}=\left(\frac{k_{B}T^{0}_{\Theta}}{\hbar \Omega}\right)S(\omega/\Omega)
\label{eq:last}
\end{equation}
The scaling function $S(\omega/\Omega)$ is constrained by the physics of the high and low frequency limits.  As $\omega/\Omega\rightarrow \infty$, $S$ must approach $i\Omega/\omega$ in order for $\sigma$ to assume its superconducting form (Eq.1).  At low frequencies $S$ approaches a real constant $Re(S(0))$ which characterizes the DC conductivity of the normal state.  
	
By comparing the measured complex conductivity to Eq. 2, we can extract both the phase stiffness and correlation time at each temperature.  To analyze the experimental data in terms of Eq. 2, we note that the phase angle of the complex conductivity, $\phi\equiv tan^{-1}(\sigma_2/\sigma_1)$, equals the phase angle of $S(\omega/\Omega)$.  Therefore $\phi$ depends only on the single parameter $\Omega$, and is independent of $T^{0}_{\Theta}$.  With the appropriate choice of $\Omega(T)$, all the measured values of $\phi$ should collapse to a single curve when plotted as a function of the normalized frequency $\omega/\Omega$.  Knowing $\Omega(T)$, $T^{0}_{\Theta}$ is obtained from a collapse of the normalized conductivity magnitude, $(\hbar\Omega/k_{B}T^{0}_{\Theta})|\sigma(\omega)|/\sigma_{Q}$ to $|S(\omega/\Omega)|$.  Fig. 3 shows the collapse of the data to the phase angle and magnitude of $S$.  As anticipated, $S$ approaches a real constant in the limit $\omega/\Omega\rightarrow 0$, and approaches $i\Omega/\omega$ as $\omega/\Omega\rightarrow \infty$.

\begin{figure}[h]
     \includegraphics[width=3in]{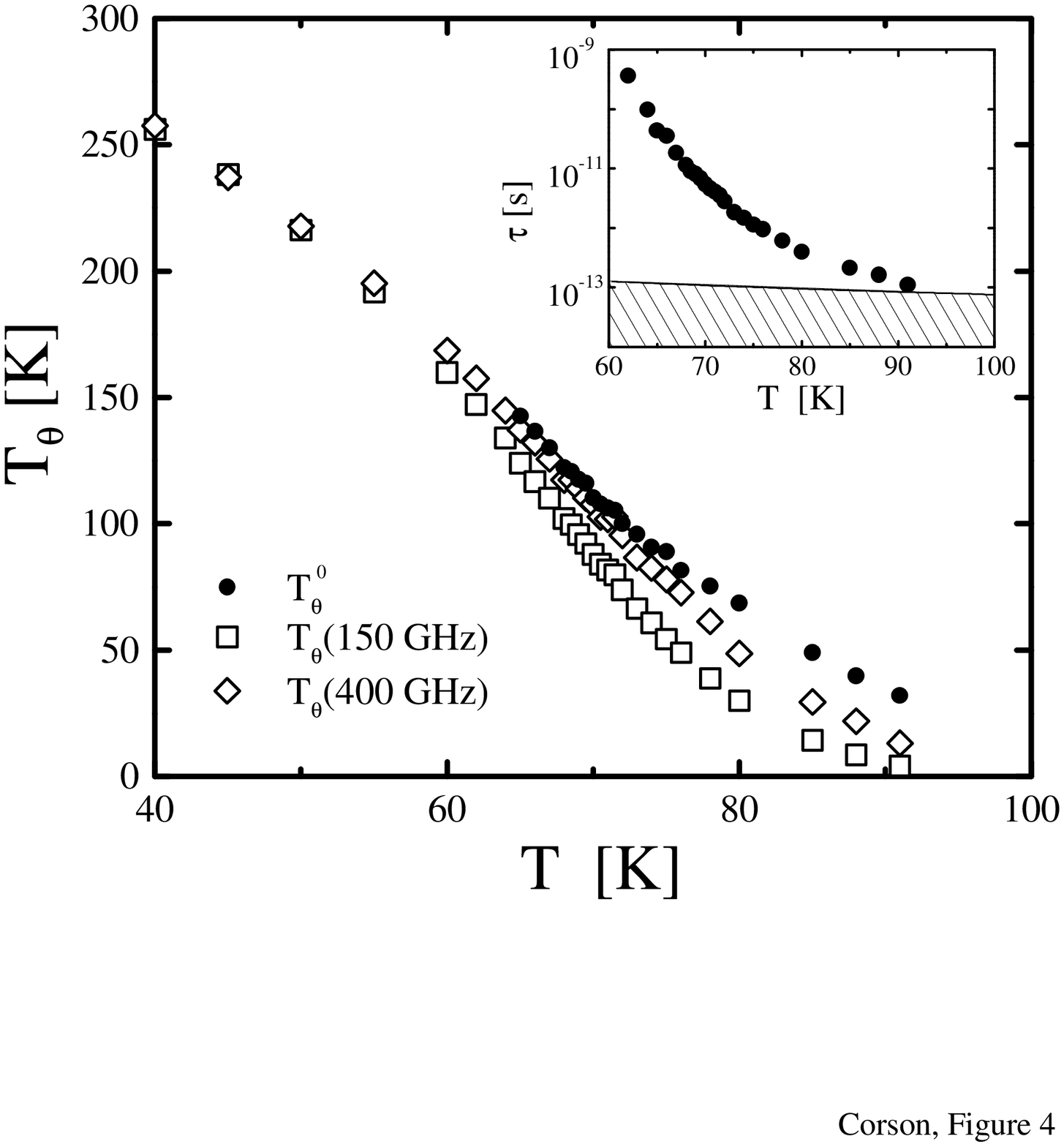}
\caption{The phase correlation time $\tau\equiv\Omega^{-1}$ and bare phase-stiffness $T^{0}_{\Theta}$ found from the scaling analysis of conductivity data are plotted as a function of temperature.  The main panel compares the bare stiffness, shown as solid circles, with the dynamic phase-stiffness at 150 GHz (open squares) and at 400 GHz (open diamonds).  The inset shows $\tau$ on a semi-log plot.  The hash marks define a region where $\tau$ is less than the lifetime of electrons in the normal state, $\hbar /k_{B}T$.  When the data points reach this region, phase fluctations of the superconducting condensate become indistinguishable from the ballistic dynamics of normal electrons.}
\label{fig:Fourth}
\end{figure}

When analyzed further, the data reveal a remarkable confirmation of thermal generation of vortices in the normal state.  In the KTB picture we expect that the DC conductivity will equal $k_{B}T/n_{f}D\Phi_{0}^{2}$, which is the "flux-flow" conductivity of $n_{f}$ free vortices with quantized flux $\Phi_{0}$, and diffusivity $D$\cite{16}.  Together with Eq. 2, this implies that $\Omega$ is a linear function of $n_{f}$, i.e., $\Omega=\Omega_{0}n_{f}a_{vc}/\Theta$, where $a_{vc}$ is the area of a vortex core, $\Theta\equiv T/T^{0}_{\Theta}$ is the reduced temperature, and $\Omega_{0}\equiv\pi^{2}Re(S(0))D/a_{vc}$.  Moreover, we expect that $n_{f}$ will be a thermally-activated function of temperature, except for $T$ very close to $T_{KTB}$.  The activation energy is simply $Ck_{B}T^{0}_{\Theta}$, where $C$ is a nonuniversal constant of order unity.  It follows that the fluctuation frequency depends exponentially on the reciprocal of the reduced temperature, $\Omega=(\Omega_{0}/\Theta)exp(-2C/\Theta)$. 

The inset to Fig. 3 is a plot of log($\Theta\Omega$) vs. $1/\Theta$ which shows that the exponential relation is observed over nearly four orders of magnitude.  This is direct evidence that vanishing of phase coherence in our samples reflects the dynamics of thermally-generated vortices.  From the slope and intercept of a straight line fit we obtain $C=2.23$ and $\Omega_{0}=1.14\times 10^{14} s^{-1}$.  

In Fig. 4 we present the behavior of the bare stiffness and phase-correlation time obtained from our measurement and modeling of $\sigma(\omega)$.  The main panel contrasts $T^{0}_{\Theta}$ with the dynamical phase stiffness $T_{\Theta}(\omega)$ measured at 150 and 400 GHz.  The inset shows $\tau$ as a function of temperature together with hashes that highlight the region where $\tau<\hbar/k_{B}T$. 

The parameters displayed in Fig. 4 suggest that while phase correlations indeed persist above $T_{c}$, they vanish well below $T^{*}$.  The loss of coherence is driven by the decrease of   $T^{0}_{\Theta}$ with increasing temperature, which renders the system increasingly defenseless to the proliferation of free vortices.  This decrease of $T^{0}_{\Theta}$ is consistent with a phenomenological description of a d-wave superconductor derived from a Mott insulator.  In this picture the phase-stiffness of an underdoped cuprate superconductor is undermined below $T^{*}$ by the thermal generation of normal electrons very near the points in momentum space where the superconducting gap vanishes\cite{17}.  The pairing which remains in other regions of momentum space appears to contribute little to the overall phase stiffness.  Near 95 K, $\tau$ falls to its minimum detectable value of $\hbar /k_{B}T$, which is the electron mean-free-time. Beyond this point superconductivity becomes indistinguishable from the ballistic dynamics of normal electrons and the recovery of the incoherent normal state is complete.  

\acknowledgments
This work was supported under NSF Grant No. FD95-10353, DOE Contract No. DE-AC03-76SF00098, and ONR Contract No. N00014-94-C-001.  J. O. thanks the Aspen Center for Physics for the opportunity to discuss this work with participants of its high-Tc workshop.

\end{document}